\documentclass[twocolumn,prb]{revtex4}
\usepackage{graphicx}
\begin{document}
\title{Mutual information and the structure of entanglement in quantum field theory}
\author{Brian Swingle}
\email{bswingle@mit.edu}
\affiliation{Department of Physics, Massachusetts Institute of Technology, Cambridge, Massachusetts 02139}
\begin{abstract}
I study the mutual information between spatial subsystems in a variety of scale invariant quantum field theories.  While it is derived from the bare entanglement entropy, the mutual information offers a more refined probe of the entanglement structure of quantum field theories because it remains finite in the continuum limit.  I argue that the mutual information has certain universal singularities that are a manifestation of the idea of ``entanglement per scale".  Moreover, I propose a method, based on an ansatz for higher dimensional twist operators, to compute the entanglement entropy, Renyi entropy, and mutual information in a general quantum field theory.  The relevance of these results to the search for renormalization group monotones, to holographic duality, and to entanglement based simulation methods for many body systems are all discussed.
\end{abstract}
\maketitle

\section{Introduction}
The many body Hilbert space is an exponentially large vector space in which the quantum state of a many body system lives.  Nevertheless, much of this space is not relevant for studying ground states of many body systems.  Recently, an exchange of ideas between quantum information science and many body quantum physics has led to an improved understanding of and appreciation for the ``corner" of Hilbert space in which quantum ground states of local Hamiltonians typically reside.  One of the earliest tools for investigating the properties of many body ground states was the entanglement entropy, defined as the von Neumann entropy of the reduced density matrix of a spatial subsystem.  The ubiquitous presence of an area or boundary law for the entanglement entropy in quantum ground states has provided a rough guide to the entanglement properties of quantum ground states \cite{arealaw1}.  This rough intuition led to a new class of quantum states generically called tensor network states \cite{mera,peps,terg} as well as new insights into the classification and identification of many body phases and phase transitions.

However, the entanglement entropy does suffer from at least one defect: it is a cutoff or high energy sensitive quantity \cite{arealaw1}.  In an unregulated quantum field theory, the entanglement entropy is formally divergent due to the presence of high energy singularities associated with the boundary law behavior.  To render the entropy finite, we must regulate the quantum field theory by providing a high energy completion such as a quantum lattice theory.  In the context of quantum field theoretic studies, therefore, the focus has been on special contributions to the entanglement entropy that can be argued to be universal.  Examples of such contributions include the logarithmic term in $1+1$ dimensional conformal field theories \cite{geo_ent,eeqft} and the topological entanglement entropy in $2+1$ dimensional topological phases \cite{topent1,topent2}. This situation is unsatisfactory: while a great deal of intuition now exists for the behavior of the entanglement entropy, such intuition is often bound up with non-universal lattice scale physics.  This is not to say that the lattice scale physics isn't of interest, only that a complete understanding of many body entanglement should contain a clear separation between the physics of low and high energies to the extent possible.

There is a quantity, called the quantum mutual information, that shares some of the features of the entanglement entropy and remains finite in a quantum field theory.  To define the mutual information, we consider two spatial subsystems, $A$ and $B$, of a larger many body system.  The mutual information is $\mathcal{I}(A,B) = S_A + S_B - S_{A \cup B}$, where $S_R$ is the entanglement entropy of region $R$.  The mutual information is positive and symmetric in $A$ and $B$ and, for separate regions, the subtraction insures that non-universal boundary law contributions cancel.  On the other hand, if we take $A$ and $B$ to form the entire system (assumed to be in a pure state) then $S_{A \cup B} = 0$ and $\mathcal{I}(A,B) = 2 S_A = 2 S_B$.  Note that this requires regions $A$ and $B$ to come together and touch, so new divergences may appear from this procedure.  At the very least, the mutual information is interesting because it captures some of the physics of entanglement, because it bounds normalized correlation functions \cite{mibound}, and because it is well defined property of a quantum field theory.

In field theoretic studies, the entanglement entropy is often computed using the replica trick: a partition function involving $n$ copies of the field theory is developed to compute the quantity $Z_R(n) = \text{tr}(\rho_R^n)$ from which the entanglement entropy of region $R$ can be obtained.  In $1+1$ dimensional conformal field theories this partition function may be computed with the help of interesting operators called twist operators \cite{eeqft,geo_ent}.  These operators turn out to be primary with conformal dimension related to the central charge of the conformal field theory.  The mutual information is a finite quantity in the conformal field theory that accesses the properties of these interesting operators.  In higher dimensions, the analog of twist operators are no longer point like, becoming instead line operators in $2+1$ dimensions and surface operators in $3+1$ dimensions.  These line and surface operators are also twist operators of a sort, and the mutual information is a finite quantity which accesses the properties of these operators.  Thus a careful study of the mutual information might reveal some information about these mysterious extended operators in higher dimensional quantum field theory.

One additional motivation for studying entanglement via the mutual information comes from the application of holographic duality to condensed matter systems \cite{maldacena,polyakov,witten,holoreview}. It is important to try to validate the new insights coming from holographic duality in the context of more ordinary quantum field theories.  I want to establish as clearly as possible exactly what physical quantities are properly captured, at least in a qualitative sense, by holographic duality even in the large N and strong coupling limits.  As a foundational element of all many body quantum systems, the structure of entanglement seems an ideal place to begin such a systematic comparison.  And there is growing evidence that holographic duality does indeed capture at least qualitatively the structure of entanglement in generic quantum field theories.

This paper is organized as follows.  I first describe the basic scaling intuition for many body entanglement.  Then I reformulate this scaling structure for the mutual information and check the proposal using holographic methods.  I also discuss in some detail a proposal for higher dimensional twist operators that permit calculation of the mutual information.  Finally, I comment on the possibility that entanglement provides a quantity that is monotonic under the renormalization group flow.

\section{Scaling intuition}
It is appropriate to begin with a brief reminder of the basic behavior of entanglement entropy in local quantum field theories.  The basic result is the ``boundary law" for entanglement entropy: for a local quantum field theory in its ground state, the entanglement entropy of a region of linear size $L$ scales as $L^{d-1}$ where $d$ is the dimension of space \cite{arealaw1}.  This rule appears to hold for theories with a gap in any dimension $d$.  It also holds for conformal field theories in more than one dimension, but for $1+1$ dimensional conformal field theories, the expected constant scaling is replaced by a term logarithmic in region size.  Such a logarithmic violation of the boundary law also appears in fermion systems with a Fermi surface \cite{fermion1,fermion2,fermion3,fermion4,fermion5}, and in fact, the two anomalies are connected \cite{bgs_ferm1,bgs_ferm2}.  As I already mentioned, however, the boundary law piece of the entanglement entropy is a cutoff sensitive quantity and is thus non-universal.  Nevertheless, there is a simple scaling picture which captures much of the variety of possible entanglement behavior.

Consider a quantum field theory in $d$ space dimensions.  I wish to analyze the behavior of entanglement as a function of energy scale.  To this end, let me introduce a variable $r$ which encodes the length scale of interest.  $r = \epsilon$ corresponds to the high energy cutoff where the field theory is superseded by some high energy completion.  $r \rightarrow \infty$ corresponds to the low energy limit of the quantum field theory, and if the field theory is gapped then this limit gives a trivial theory.  As in standard renormalization group treatments, $r$ changes as we move along the renormalization group flow, and the appropriate measure for changes in $r$ is the logarithmic measure $dr/r$.

I wish to make a hypothesis about the entanglement at a scale $r$ in order to recover the familiar boundary law scaling of entanglement entropy.  To motivate the coming assumption, consider the coarse grained Hamiltonian $H(r)$ as a function of $r$.  At each scale $r$, this Hamiltonian is local at scales longer than $r$, for example, the microscopic Hamiltonian $H(\epsilon)$ is local at the lattice scale.  Thus, considering a region of size $L$ and its complement, the coarse grained Hamiltonian at scale $r$ only entangles the region with its environment along the boundary of the region.  Now, the rough number of degrees of freedom at the boundary of the region of size $L$ at scale $r$ is $\left(\frac{L}{r}\right)^{d-1}$ in $d$ space dimensions.  If I assume that each degree of freedom contributes roughly one ``ebit" to the entanglement entropy, then the contribution to the entanglement entropy at scale $r$ is
\begin{equation}
dS(r) = \left(\frac{L}{r}\right)^{d-1} \frac{dr}{r}.
\end{equation}

To obtain the full entanglement entropy I simply integrate this contribution from the high energy cutoff down to an appropriate low energy cutoff:
\begin{equation}
S = \int^{r_{IR}}_{r_{UV}}  \left(\frac{L}{r}\right)^{d-1} \frac{dr}{r}.
\end{equation}
The high energy or UV cutoff is simply $r_{UV} = \epsilon$, but the low energy or infrared cutoff depends on the nature of the theory.  For a conformal field theory, the only scale is the region size $L$, so the infrared cutoff is the region size $r_{IR} = L$.  This naturally reproduces the boundary law in dimension $d >1$ and the logarithmic violation in $d=1$.  I would also like to point out that this scaling ansatz also naturally shows that corners in a conformal field theory can be associated with logarithmic corrections because corners can contribute a fixed amount of entanglement at every scale giving $\int dr/r \sim \log{L}$.  On the other hand, if the theory has a finite correlation length, then the infrared cutoff is given by $r_{IR} = \min{(L,\xi)}$ where $\xi$ is the correlation length.  In this case one always obtains a boundary law for sufficiently large $L$.  Note that the ``entanglement per scale" in the one dimensional conformal case is a quantity of some interest, namely the central charge of the conformal field theory.  I would like to say that in any conformal field theory in any dimension, the notion of ``entanglement per scale" is a well defined and universal quantity.  However, the entanglement entropy as it stands is bound up with non-universal cutoff scale physics and cannot provide a clean definition of ``entanglement per scale".

To find a suitable formalism for extracting the physics of entanglement as a function of scale, I turn to the quantum mutual information.  The mutual information between two regions $A$ and $B$ is defined as $\mathcal{I}(A,B) = S_A + S_B - S_{A \cup B}$.  Subadditivity of the von Neumann entropy guarantees that the mutual information is a positive quantity, and the mutual information is manifestly symmetric in $A$ and $B$.  It measures in a uniform way the degree of correlation between regions $A$ and $B$.  If the density matrix $\rho_{AB}$ factorizes into $\rho_A \otimes \rho_B$ then the mutual information vanishes.  The converse is also true.  The mutual information also gives more than just a yes/no answer to the question of correlations: it bounds the connected correlation functions of operators localized in $A$ and $B$.  In particular, we have $\langle \mathcal{O}_A \mathcal{O}_B \rangle_c^2 \leq ||\mathcal{O}_A||^2 ||\mathcal{O}_B||^2 \mathcal{I}(A,B)$ \cite{mibound}.

The crucial property of the mutual information that makes it useful for my purposes is its cutoff independence.  Indeed, the boundary law terms containing information about physics of the cutoff cancel in the subtraction that defines the mutual information. From another point of view, this independence from the cutoff can arise because we necessarily introduce additional length scales when considering multiple regions.  For example, the distance between regions provides an additional scale beyond the size of each region.  Let me now turn to reformulating the scaling intuition described above in terms of the mutual information.

\section{Universal singularities}
\subsection{$1+1$ dimensions}
I begin with the case of one dimensional conformal field theory, in particular, the case of free fermions.  In general, knowledge of the entanglement entropy for multiple regions, as required to compute $\mathcal{I}(A,B)$, is not trivially related to the single region result.  The single region entanglement entropy in any conformal field theory in one dimension contains a universal logarithmic term depending only on the central charge. but the multi-region entanglement entropy is known to depend on the entire field content of the conformal field theory.  However, for free fermions the result is actually known even for multiple intervals \cite{ee1df}.  We will use only the result for two intervals specified by $[a_i, b_i]$ with $i=1,2$.  The entanglement entropy is
\begin{widetext}
\begin{equation}
S_{\text{2 region}} = \frac{1}{3} \left(\sum_{i j} \log{\left(\frac{|a_i - b_j|}{\epsilon}\right)} - \log{\left(\frac{|a_1 - a_2|}{\epsilon}\right)}  -\log{\left(\frac{|b_1 - b_2|}{\epsilon}\right)}\right).
\end{equation}
\end{widetext}

For simplicity I consider the case of two intervals of equal size $L$ separated by a distance $x$ defined as the nearest distance between the two lines.  The two interval entanglement entropy becomes
\begin{widetext}
\begin{equation}
S_{\text{2 region}} = \frac{1}{3}\left( 2 \log{\left(\frac{L}{\epsilon}\right)} + \log{\left(\frac{x}{\epsilon}\right)} + \log{\left(\frac{2L+x}{\epsilon}\right)} - 2 \log{\left(\frac{L+x}{\epsilon}\right)}\right).
\end{equation}
\end{widetext}
To obtain the mutual information between the two intervals, the two interval result is subtracted from the sum of the entanglement entropy of each region separately:
\begin{equation}
\mathcal{I}(L,x) = \frac{1}{3}\left(2 \log{\left(\frac{L}{\epsilon}\right)}\right) - S_{\text{2 region}}.
\end{equation}
There are many cancelations in this equation, and in particular, the cutoff dependence completely disappears as promised.  The final result is
\begin{equation}
\mathcal{I}(L,x) = \frac{1}{3} \log{\left(\frac{(L+x)^2}{x(2L+x)}\right)}.
\end{equation}

Something remarkable happens as as $x$ goes to zero, that is, as the two regions approach each other: the mutual information contains a universal divergence going as $\log{x}$.  Moreover, the coefficient of this divergence is precisely the central charge of the free fermion CFT that we wanted to interpret as the entanglement per scale in a conformal field theory.  Now I state a more general result: in any conformal field theory, the leading singularity in the mutual information as two regions approach each other is universal and given by the central charge of the conformal field theory \cite{eeqft}.  Thus, despite the complicated nature of the mutual information in general, the singularity structure as regions collide is highly constrained.  This result follows from the short distance properties of the twist operators that define the entanglement entropy in the replica version of the original CFT.  These twist fields are primary with a dimension related to the central charge of the original CFT, and the leading term in their OPE is a fusion to the identity.

\subsection{Higher dimensions}

I will return to the subject of twist operators later, but for now, let me try to generalize this result to CFTs in higher dimensions. I will not try to directly compute the full mutual information in a conformal field theory in higher dimensions.  Although this calculation may be possible in some cases, I am in this section only interested in certain universal divergences that appear as regions are brought together.  Now, a natural question in higher dimensions, which does not arise in one dimension, is the precise nature of this collision process.  There are several ways in which one can imagine performing this procedure.  First, if regions $A$ and $B$ have a flat $d-1$ dimensional surface, then we can bring the regions together along this surface.  More generically, if the boundaries of the two regions are smooth, they will typically only touch at a single point with finite radius of curvature.  A final interesting possibility is the case of sharp corners approaching each other or another smooth interface.  I will address all three situations below.

Consider first the case of a collision of flat $d-1$ dimensional surfaces.  I want to know how the mutual information behaves as a function of $x$, the separation between the two flat sections of the boundary.  The flat sections are taken to have equal $d-1$ dimensional size $V_{d-1}$.  Based on the boundary law for entanglement entropy, and because I expect to recover part of the divergent entanglement entropy when the regions touch, the mutual information should diverge as $x \rightarrow 0$.  By analogy with the boundary law, I find that
\begin{equation}
\mathcal{I}(x) = k \frac{V_{d-1}}{x^{d-1}} + ...,
\end{equation}
where $...$ indicates subleading terms in $1/x$.  The coefficient $k$ in this expression should be a universal quantity that effectively counts the number of degrees of freedom in the conformal field theory (in terms of how much entanglement they contribute to the ground state).  In a sense, this is the boundary law but with the non-universal cutoff $\epsilon$ replaced by a definite continuum quantity $x$, the separation between the $d-1$ dimensional surfaces.  Indeed, if we assume that the divergence must be proportional to the size of the colliding regions then the dependence on $x$ is fixed by dimensional analysis up to logarithmic corrections.  Of course, such a logarithmic correction is realized in one dimension where we already found the result $\mathcal{I}(x) \sim \log{x}$, and there the constant $k$ is proportional to the central charge.

But what happens in the more generic situation where the regions $A$ and $B$ collide only at a single point.  The mutual information will still diverge, but with a weaker power of $x$.  Suppose the two regions touch at a single point and that in the neighborhood of each point the boundary may be described as a parabolic surface rotationally invariant about the $x$ axis with radius of curvature $R$.  The mutual information should be a universal function of $R/x$.  I will now give a scaling argument to determine this function.  Consider two parabolic surfaces separated by a distance $x$ parameterized by a radial coordinate $\rho$.  The length of the line parallel to the $x$ axis connecting the two surfaces at radial coordinate $r$ is $x + \frac{\rho^2}{R}$.  Let us now apply the form deduced above for the scaling form of the mutual information in the case of flat regions to a small shell with inner radius $r$ and outer radius $r+dr$.  The approximate size $V_{d-1}$ of this shell is $\rho^{d-2} d\rho$, and the distance between the shells along the $x$ axis is $x+\frac{\rho^2}{R}$.  I now integrate the mutual information obtained above from $\rho=0$ to some cutoff value $\rho_c$:
\begin{equation}
\mathcal{I}(R,x) \sim \int^{\rho_c}_0 d\rho \frac{\rho^{d-2}}{(x + \rho^2/R)^{d-1}}.
\end{equation}
This formula is divergent as $x$ goes to zero, and to determine the scaling form we simply $\rho = \sqrt{Rx} u$ to find
\begin{equation}
\mathcal{I} \sim \int^{u_c}_0 \sqrt{R x} du \frac{(R x)^{d/2 -1}u^{d-2}}{x^{d-1} (1+u^2)^{d-1}}.
\end{equation}
Collecting all the powers of $x$ and $R$, I find that
\begin{equation}
\mathcal{I}(R,x) = k' \left(\frac{R}{x}\right)^{\frac{d-1}{2}} + ...
\end{equation}
Thus, the mutual information still diverges but with a different power of $x$ owing to the quadratic nature of the boundaries near the collision point.

Finally, one can consider the singular situation of a corner approaching a smooth surface.  This case is similar to that of colliding quadratic surfaces except that the distance along the $x$ axis between shells depends linearly on $\rho$.  Carrying out the same integral as above with $x+\rho^2/R$ replaced by a linear function $x + m\rho$ ($m$ is a function of the opening angle) gives a logarithmic divergence $\mathcal{I} = k'' \log{x} + ...$ in any dimension $d$.  Not only do $d-1$ dimensional flat surfaces replicate the boundary law, colliding corners also replicate the logarithmic term in the entanglement entropy associated with corners in any dimension.

I have argued on general grounds that the mutual information in a conformal field theory should contain certain universal divergent pieces when the regions involved collide.  Depending on the geometry of the collision, one obtains different scaling forms with universal prefactors that measure the number of degrees of freedom contributing to ground state entanglement.  However, it is desirable to check these scaling relations in specific cases to explore the validity of the arguments just given.  Thus, I now turn to class of theories for which the above conjectured scaling forms can be explicitly verified.

\section{Holographic computation}
The simplest setting in which these ideas can be tested is provided by holographic duality.  Holographic duality relates quantum field theories in $d+1$ spacetime dimensions to theories of quantum gravity in curved higher dimensional spaces \cite{maldacena,polyakov,witten,holoreview}.  The classic statement of the duality is between $\mathcal{N}=4$ super Yang-Mills theory in four dimensions and type $IIB$ string theory on asymptotically $\text{AdS}_5 \times S^5$ spacetimes.  The high energy limit of the field theory is in some sense located at the conformal boundary of $\text{AdS}_5$, and the extra radial dimension of $\text{AdS}_5$ is associated with energy scale in the field theory.  The duality becomes particularly simple on the gravity side when we take the limit of large $N$ and large $\lambda = g^2_{YM} N$ in the field theory.  In this limit, the string theory becomes well approximated by classical (coming from large $N$) supergravity (coming from large $\lambda$).  Many interesting quantities in this strong coupling limit of the field theory become expressible as simple geometric objects in a higher dimensional spacetime with gravity.

\begin{figure}
\includegraphics[width=.48\textwidth]{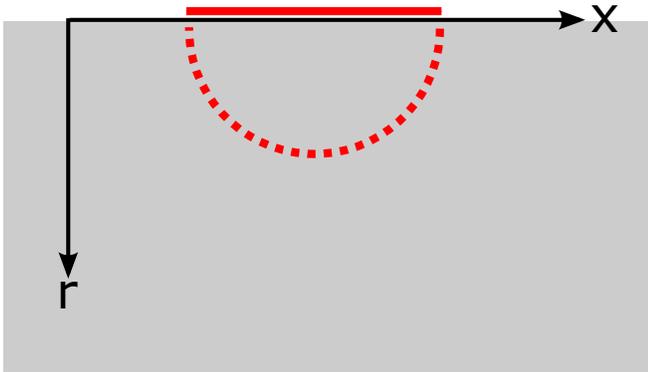}
\caption{Sketch of the prescription for computing the entanglement entropy in a $1+1$ dimensional CFT via holographic duality.  The horizontal axis is the spatial coordinate in the CFT while the $r-$axis is the bulk radial coordinate (the bulk is shown in gray).  The time coordinate is suppressed.  The region of interest in the field theory is shown as a thick red bar while the minimal surface of interest in the bulk is the red dotted line.  The minimal surface is required to terminate at the boundary of the region in the field theory.}
\end{figure}

In particular, the entanglement entropy in the field theory is related to minimal surfaces in the higher dimensional gravitational geometry \cite{holo_ee,holo_ee_f}.  The detailed prescription is as follows.  As we said, the high energy limit of the field theory lives at the conformal boundary of AdS.  To compute the entanglement entropy holographically, we must study surfaces in the bulk gravitational geometry that asymptote at the conformal boundary of AdS to the boundary of the region in the field theory we are interested in.  The entanglement entropy is then the area in units of the Planck length of the minimal area surface satisfying the boundary conditions.  The prescription is illustrated in Fig. 1 for the case of a $1+1$ dimensional conformal field theory dual to an AdS$_3$ geometry.  Using this prescription we can reduce the computation of the mutual information in holographic theories to a certain minimization problem in a curved higher dimensional geometry.

\begin{figure}
\includegraphics[width=.48\textwidth]{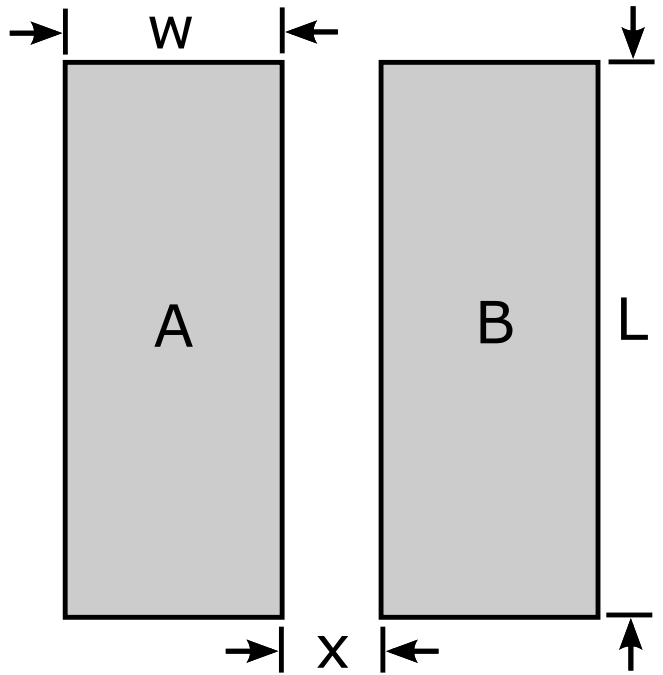}
\caption{Strip geometry for the calculation of mutual information in the translation invariant case.  We assume the $L \gg w \gg x$ so that the minimal surface problem reduces to a single variable problem.  A singularity in the mutual information develops as $x$ approaches zero.}
\end{figure}

Let us first consider the case of colliding $d-1$ dimensional surfaces in a $d+1$ dimensional conformal field theory ($d$ is the dimension of space).  I will focus on the case of $d=2$ for ease of presentation, but the results are quite general.  The metric of AdS$_{3+1}$ is
\begin{equation}
ds^2 = \frac{L_{\Lambda}^2}{r^2}\left(-dt^2 + dr^2 + dx^2 + dy^2\right),
\end{equation}
where $L_{\Lambda}$ is the AdS radius and $r$ is the radial coordinate.  Let the field theory regions $A$ and $B$ be strips of length $L$ in the $y$ direction with width $w$ in the $x$ direction.  The strips are assumed to be separated by a distance $x$ as illustrated in Fig. 2.  Assuming $L \gg w$, we have translation invariance in the $y$ direction.  This greatly simplifies the minimal surface problem, allowing us to parameterize the minimal surface by $r(x)$ independent of $y$.  Focusing first on a single strip, the area of the surface in the bulk is
\begin{equation}
L_{\Lambda}^2 L \int^{w/2}_{-w/2} dx \frac{\sqrt{1+\left(\frac{dr}{dx}\right)^2}}{r^2}.
\end{equation}
The result of the minimization procedure is an area of the form
\begin{equation}
L_{\Lambda}^2 \left(k_1 \frac{L}{\epsilon} - k_2 \frac{L}{w}\right),
\end{equation}
where the constants $k_1$ and $k_2$ are calculable and $\epsilon$ is a high energy cutoff.  The entanglement entropy of the single strip is simply this area multiplied by $1/(4 G_N)$, that is, the area in Planck units.

As usual, the entanglement entropy is non-universal, depending on the high energy cutoff $\epsilon$.  To remove this defect we return now to the two strip geometry and compute the mutual information for the two strips.  We need two quantities: the entropy of a single strip and the entropy of both strips together.  We have already obtained the single strip entropy, so let us focus on the two strip problem.  There are two cases that must be considered depending on the ratio of the strip width and the strip separation.  These two cases correspond to two possible choices for the two strip minimal surface.  The first choice corresponds to two disconnected surfaces, one for each strip, and each identical to the single strip minimal surface.  This situation occurs when the strips are widely separated and gives zero mutual information.  However, for the purposes of calculating universal divergences in the mutual information we are interested in the opposite limit of two very close strips.  In this case, the minimal surface actually connects the two strips in the bulk.  The near edges of the strips are connected by one component of the bulk minimal surface while the far edges are connected by another component.  This geometry is illustrated in Fig. 3.

\begin{figure}
\includegraphics[width=.48\textwidth]{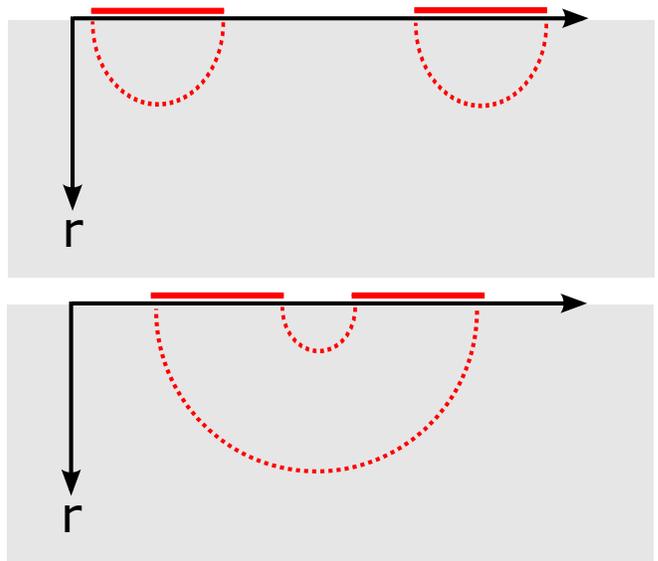}
\caption{A sketch of the two bulk minimal surfaces relevant for the calculation of the mutual information.  The translation invariant spatial coordinate is suppressed along with the time.  In the top panel, the minimal surface for two widely separated strips is simply two copies of the minimal surface for a single strip.  In the bottom panel, when the two regions come close, a new minimal surface appears which connects the inner and outer boundaries of the two regions.  In this case, there is a non-zero holographic mutual information.}
\end{figure}

Repeating the analysis above for the minimal surface, I find that the entanglement entropy of the two strips taken together is
\begin{equation}
S_{\text{2 strips}} = \frac{L_{\Lambda}^2}{4 G_N}\left( \left(k_1 \frac{L}{\epsilon} - k_2 \frac{L}{2w+x}\right)+ \left(k_1 \frac{L}{\epsilon} - k_2 \frac{L}{x}\right)\right).
\end{equation}
Now, the mutual information is $\mathcal{I} = 2 S_{\text{1 strip}} - S_{\text{2 strips}}$ which gives
\begin{equation}
\mathcal{I}(L,w,x) = k_2 \frac{L_{\Lambda}^2}{4 G_N} \left(\frac{L}{x} + \frac{L}{2w+x} - 2\frac{L}{w}\right).
\end{equation}
Note again that the mutual information is manifestly cutoff independent.  The factor $L_{\Lambda}^2/G_N$ is related to the total number of local degrees of freedom in the field theory, for example, it may be related to the dimension of the gauge group ($\sim N^2$ for SU($N$)).  As promised, the mutual information has a universal divergence as $x\rightarrow 0$, and this divergence is proportional to the size of the colliding region, here the length $L$, and to the total number of degrees of freedom.

We can also ask what happens when the colliding regions are not flat surfaces but points with finite radius of curvature or corners.  Of course, the minimal surfaces in this case which connect the two regions will be much more complicated, but I will argue that we do not need the full minimal surface to verify the scaling form proposed above.  Consider first the case of two long strips, but now with the colliding side of each strip curved into a portion of a circle with a very large radius of curvature $L$ as shown in Fig. 4.  Now the two regions collide only at a single point, but nevertheless, the minimal surface will be approximately translation invariant in the $y$ direction.  Let us now parameterize the minimal surface as $r = r(x,y)$ but with the expectation that $\partial_y r \ll \partial_x r$ except possibly deep in the bulk, in other words, we suppose the surface varies slowly with $y$.  The full expression for the minimal surface area is
\begin{equation}
L_{\Lambda}^2 \int dx dy \frac{\sqrt{1+(\partial_x r)^2 + (\partial_y r)^2}}{r^2},
\end{equation}
and we want to be in a regime where $\partial_y r \ll \sqrt{1+(\partial_x r)^2}$ for all $x$ and $y$.  In this regime, there is a separation of scales between the fast variable $x$ and the slow variable $y$ and we can approximately solve the fast problem treating the slow variable as fixed.  This means we obtain the same sort of minimal surface as we found above for the infinite strip except that the strip width is a local quantity determine by the slow variable $y$.  Only in the case where the strip width was independent of $y$ were we able to perform the $y$ integral exactly to yield $L$, the strip length.

\begin{figure}
\includegraphics[width=.48\textwidth]{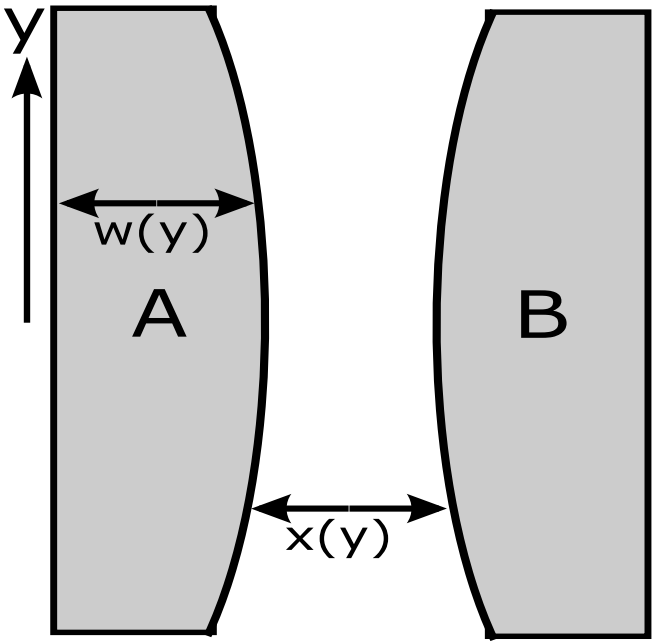}
\caption{The analog of the strip geometry, but with one flat side of each strip replaced by an arc of a circle with large radius of curvature.  The resulting regions are now nearly translation invariant in the $y$ direction, but also only collide at a point as $x(y)$ approaches $0$.}
\end{figure}

Thus a segment $y$ to $y+dy$ contributes an infinitesimal mutual information given approximately by
\begin{equation}
d\mathcal{I}(w(y),x(y)) = k_2 \frac{L_{\Lambda}^2}{4 G_N} \left(\frac{dy}{x} + \frac{dy}{2w+x} - 2\frac{dy}{w}\right),
\end{equation}
where $x(y)$ and $w(y)$ are the width and separation of the slightly curved strips as a function of $y$, the coordinate along the long length of the curved strips.  In the limit of large radius of curvature $L$ we may approximate the width as constant, $w(y) \approx w_0$, while the separation is approximated by a quadratic function, $x(y) \approx x_0 + y^2/L$, so that only the point $y=0$ at the middle of the curved strip actually collides in the limit $x_0 \rightarrow 0$.  Focusing now on the leading divergence in the mutual information, we find that we must determine the singularity in the integral $\int dy \frac{1}{x_0 + y^2/L}$ as $x_0$ goes to zero, but this is precisely the integral we considered above where we found that it diverges as $\sqrt{L/x_0}$ as $x_0$ goes to zero.  Thus the holographic prescription for entanglement entropy reproduces the intuitive scaling we argued for above, at least in the nearly translation invariant limit.  A completely analogous argument also shows that corners in the holographic case give a logarithmic singularity $\log{x_0}$ simply because the separation $x(y)$ is in that case a linear function of $y$.

All the explicit computations up to this point have been for a holographic field theory with $2$ spatial dimensions.  Of course, we can generalize these results to higher dimensions.  The geometry is more complicated, but the results are unchanged, so I do not include the details here.  Note also that we relied entirely on the spatial geometry of AdS for these computations, and this implies that the scaling we obtained for the universal divergences in the mutual information are identical in form for holographic $z \neq 1$ scale invariant theories with spatial AdS slices \cite{holonr1,holonr2}.  There is also a connection between the minimal surface calculations described here and the structure of the multiscale entanglement renormalization ansatz (MERA), a class of variation quantum states \cite{ent_ren_holo}.  The minimal surfaces described here all have an interpretation on the MERA side in terms of the number of disentanglers cut by the tracing procedure, and in particular, the crossover between the short distance and long distance behavior of the mutual information is evident in terms of whether the regions entering the mutual information are renormalized to the lattice scale before or after they are merged together in the MERA.

\section{Fermi liquids}
So far I have considered mostly relativistic conformal field theories in any dimension, although the holographic results above also applied to non-relativistic scale invariant theories with dynamical exponent $z\neq 1$.  However, once one is willing to consider non-relativistic situations, there are a number of interesting renormalization group fixed points to investigate.  The simplest such fixed point (really fixed manifold) is the Fermi liquid fixed point in $d>1$ spatial dimensions (in $d=1$ we have the usual Luttinger liquid fixed line).  This fixed point is applicable for fermions at finite density with short range interactions and is characterized by scaling towards a surface, the Fermi surface, in momentum space, rather than scaling towards a single point in momentum space \cite{fermion_rg1,fermion_rg2,fermion_rg3}.  The Fermi liquid fixed point is quite interesting for my purposes because its entanglement structure is controlled by the $1+1$ dimensional nature of the ``radial" excitation near the Fermi surface \cite{bgs_ferm1,bgs_ferm2,renyi_ff}.

Thus the result for two colliding $d-1$ dimensional surfaces of size $V_{d-1}$ in a Fermi liquid differs from the $d+1$ dimensional conformal case.  In fact, it resembles the $1+1$ dimensional conformal result because the Fermi surface can be thought of as a collection of $1+1$ dimensional conformal field theories, namely the local radial fermionic excitations which propogate with Fermi velocity normal to the Fermi surface.  The result for the mutual information, using the prescription given in \cite{renyi_ff}, is
\begin{equation}
I \sim k_F^{d-1} V_{d-1} \log{x} + ...
\end{equation}
Note that this result must be interpreted somewhat carefully because of the presence of the extra scale $k_F$.  Indeed, if $x \ll k_F^{-1}$ then the mutual information will begin to probe the higher energy theory from which the Fermi liquid descends, perhaps a lattice theory or some relativistic conformal field theory perturbed by a finite chemical potential.  Thus there is a scaling regime where $x$ is small but not so small that $k_F x \sim 1$, and in this scaling regime the dominant term in the mutual information does behave like $\log{x}$.  I would also point out that if there are other gapped bosonic modes in the Fermi liquid then the mutual information will also contain contributions from these modes, and such a mode contributes a divergence of the type described above for ordinary scale invariant theories once the separation $x$ becomes less than the correlation length $\xi$ (or inverse mass, in the relativstic language) of the bosonic mode.

Of course, this is a relatively weak singularity.  In fact, the entanglement structure is such that the mutual information only diverges if flat $d-1$ dimensional segments collide.  The case of a point with finite radius of curvature colliding produces only a non-divergent cusp-like behavior in the mutual information.  This result follows from the prescription for the mutual information given in \cite{renyi_ff} upon taking into account the geometry of the colliding regions.  I expect similar behavior for the mutual information in a critical Fermi surface \cite{crit_fs}, although there is a complication mentioned above of additional massless degrees of freedom, perhaps a gauge field, a particular angular momentum channel density fluctuation, or a $z =\infty$ low energy CFT \cite{fs_boson1,fs_boson2,nfl1}.

\section{Higher dimensional twist operators}
\subsection{Definition of twist operators}
I have now described the basic scaling intuition for the mutual information and confirmed this intuition in the framework of holographic duality.  Additionally, I have described how the story changes for other kinds of non-relativistic fixed points such as Fermi liquids.  I am now ready to discuss the concept of higher dimensional operators, but first a review the situation in one dimension in appropriate.  Twist operators are often invoked in the calculation of entanglement entropy in one dimensional conformal field theory where the entanglement entropy is written as a path integral on a multi-sheeted Riemann surface via the replica trick:
\begin{equation}
S_R = \lim_{n\rightarrow 1}- \partial_n \text{tr}(\rho_R^n).
\end{equation}
Let me first consider the case of a single interval.  The path integral of the original $1+1$ dimensional conformal field theory on this multi-sheeted surface is traded for a path integral in a new conformal field theory, the symmetric product of $n$ copies of the original conformal field theory.  The relevant path integral in this symmetric product CFT is not quite the free path integral, however, as certain point-like fields called twist operators must be inserted at the two boundary points of the interval for which we are interested in the entanglement entropy \cite{eeqft}.  These twist fields account for the conical singularity that was present in the original multi-sheeted surface formulation.

For a field $\Psi$ in the original CFT, let $\Psi_{\alpha}$ denote the $n$ copies of $\Psi$ in the symmetric product theory.  The role of the twist operators is to produce a shift $\Psi_{\alpha} \rightarrow \Psi_{\alpha \pm 1}$ as a field $\Psi_{\alpha}$ encircles the twist operator in spacetime.  This shifting operation is the analog in the original multi-sheeted formulation of moving from one sheet $\Psi_{\alpha}$ to the next $\Psi_{\alpha + 1}$.  One may compute the entanglement entropy in the original conformal field theory in terms of correlation functions of these twist operators.  Thus it is valuable to know the properties of these operators for the purposes of computing entanglement entropy, and vice versa, a knowledge of the entanglement entropy for general regions provides a handle on the properties of these operators.

Having argued that the quantum mutual information contains certain universal singularities, I would like to know how to translate this into the language of higher dimensional twist operators.  The twist operators in $1+1$ dimensional conformal field theory are actually primary operators with dimension related to the central charge of the CFT and the number of replica fields $n$.  What is the analog of this statement in higher dimensions?  Before I investigate the properties of higher dimensional twist operators, I must attempt to give a clearer definition of these operators.  As in the $1+1$ dimensional case, the entanglement entropy is related to a path integral over a multi-sheeted higher dimensional spacetime as illustrated in Fig. 5.  The spacetime looks locally unexceptional except at the boundary of the region for which one is computing the entropy, and on this boundary, there is a conical singularity in the spacetime associated with the joining of the $n$ copies of the path integral.  Like in the $1+1$ dimensional case, where the boundary of a set of intervals was a set of points having codimension $2$ in spacetime, in higher dimensions the boundary also has codimension $2$ in spacetime.

\begin{figure}
\includegraphics[width=.48\textwidth]{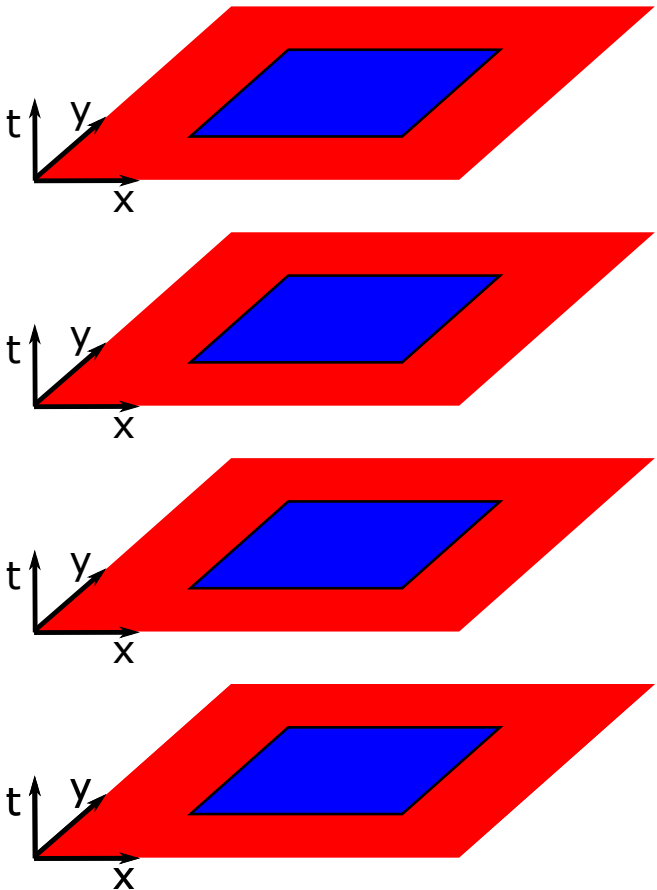}
\caption{An example of the replica method with $n=4$ copies.  The region whose entanglement entropy we are calculating is in blue while the rest of the system is in red.  The $t$ axis is imaginary time.  The copies are glued together so that one passes from copy $\alpha$ to copy $\alpha+1$ when passing through $t=0$ from below in a blue region, while in the red regions no such transition occurs.  The blue region at $t=0$ is thus a ``branch surface" that terminates on a spacetime codimension $2$ conical singularity given by the boundary between red and blue at $t=0$.  The twist operator lies along this $1$ dimensional locus in spacetime.}
\end{figure}

To formalize these notions, consider a region $\mathcal{R}$ with boundary $\partial \mathcal{R}$ in $d$ spatial dimensions.  I will focus on the case of a relativistic conformal field theory, but my considerations are generalizable, for example, to Fermi liquids.  Following the usual replica trick methods, the entanglement entropy of this region is related to a multi-sheeted path integral with a conical singularity along the boundary $\partial \mathcal{R}   $.  Alternatively, we may define an operator $K_n[\mathcal{R}]$ in the $n$-fold symmetric product theory by the equation $\langle K_n[\mathcal{R}] \rangle_n = \text{tr}(\rho_{\mathcal{R}}^n)$.  It follows from the definition that in the limit $n\rightarrow 1$ and in the absence of other operator insertions, the operator $K_n$ becomes trivial since $\text{tr}(\rho_{\mathcal{R}}) = 1$.  I also assume that this operator is localized along the boundary $\partial \mathcal{R}$ of the region $\mathcal{R}$ at a fixed imaginary time.

\subsection{Twist field ansatz}
I want to make a guess as to the form this operator by analogy with the one dimensional conformal case.  The key realization is that the twist field in $1+1$ dimensional CFT shares many properties with the exponential of a massless field, in other words, it behaves much like a vertex operator in a free boson CFT.  Thus, let us assume that the twist field in higher dimensions can also be thought of as an exponential of a massless field of some type.  Let us make the following ansatz for the form of the twist field:
\begin{equation}
K_n[\mathcal{R}] = \exp{\left(i \alpha_n \int_{\partial \mathcal{R}} \phi^{(d-1)} \right)},
\end{equation}
where $\phi^{(d-1)}$ is taken to be a massless spatial $(d-1)$-form field of scaling dimension $d-1$.  Alternatively, one may trade this $(d-1)$-form for a spatial vector using the fixed background metric
\begin{equation}
K_n[\mathcal{R}] = \exp{\left(i \alpha_n \int_{\partial \mathcal{R}} \hat{n} \cdot \vec{\phi} \right)},
\end{equation}
where $\hat{n}$ is the unit normal to the boundary $\partial \mathcal{R}$.  A knowledge of the correlators of $\phi$ in a given conformal field theory would be sufficient to compute the entanglement entropy for any region.

My ansatz for the twist fields can already reproduce all the scaling features discussed above under the particular assumption that $\phi$ has Gaussian correlations.  Of course, this assumption cannot be correct in most cases, but it does capture the short distance singularity structure nicely.  This is not so unreasonable, since the short distance structure in $1+1$ dimensions also depended on the two point function.  What follows is a sketch of the structure of these twist operators, but it is certainly not the complete story and much remains to be understood.  For example, possible complications due to phase transitions in the replicated theory as a function of $n$ are not captured in the sketch below \cite{eevect}.

To begin, observe that the field $\phi^i$ ($i=1$ to $d$ is a spatial vector index) has a shift symmetry $\phi^i \rightarrow \phi^i+a^i$ for any constant vector $a^i$.  The twist operators $K_n$ are invariant under this symmetry because the boundary $\partial \mathcal{R}$ is closed so that $\int \hat{n}\cdot \vec{a} = 0$.  This shift symmmetry is a specific example of a more general symmetry, namely the ability to shift $\phi^i$ by any vector field with zero divergence.  In the form language, this is the statement that the form field $\phi^{(d-1)}$ has a ``gauge symmetry" $\phi^{(d-1)} \rightarrow \phi^{(d-1)} + d f^{(d-2)}$ with $f^{(d-2)}$ an arbitrary smooth spatial $d-2$ form.  There is clearly a strong analogy between these twist operators and the Wilson and 't Hooft lines of gauge theories, or more generally, between the twist operators and surface operators in $p$ form gauge theories.

As Gaussian massless fields of dimension $d-1$, the correlation functions of the $\phi^i$ are determined by the basic two point function
\begin{equation}
\langle \phi^i(x) \phi^j(0) \rangle = \frac{\delta^{i j} + ...}{|x|^{2(d-1)}},
\end{equation}
where $...$ denotes additional terms like $\hat{x}^i \hat{x}^j$ which are not essential for our purposes.  As a warmup, let me compute the entanglement entropy of a circular region in $d=2$ spatial dimensions using the twist field.  One must evaluate $\langle K_n \rangle$, but the assumption of Gaussian correlations for $\phi$ gives
\begin{equation}
\langle K_n \rangle = \exp{\left(-\frac{\alpha_n^2}{2}\int_{\partial\mathcal{R}}\int_{\partial \mathcal{R}} \hat{n}^i \hat{n}^j \langle \phi_i \phi_j \rangle\right)}.
\end{equation}
What is the meaning of this double integral over the circular boundary in $d=2$?  First, it is certainly divergent and depends on the cutoff, but this is exactly what I expect for the bare entanglement entropy.  Second, I must assume that while the limit of $\alpha_n$ as $n$ goes to $1$ is zero, the combination $\alpha_n^2$ has a finite first derivative at $n=1$.  This is reasonable in light of one dimensional conformal field theory.  In that case, $\alpha_n$ is roughly the square root of the dimension $\Delta_n$ of the twist field, and $\Delta_n$ does indeed have a non-zero derivative at $n=1$.

From $\langle K_n \rangle $ one can compute the entanglement entropy via $S = -\partial_n \text{tr}(\rho^n)|_{n=1} = -\partial_n \langle K_n \rangle|_{n=1}$.  I find
\begin{equation}
S = \frac{\partial_n (\alpha^2_n)|_{n=1}}{2}  \int_{\partial\mathcal{R}}\int_{\partial \mathcal{R}} \hat{n}^i \hat{n}^j \langle \phi_i \phi_j \rangle.
\end{equation}
To evaluate the divergent double integral over $\partial \mathcal{R}$ I use the fact that no special point is selected on the circle.  Take the normal $\hat{n}_1$ of the first copy of the circle to point vertically and call $\theta$ the angle between the vertical and second normal $\hat{n}_2$.  This gives the distance $|x| = 2 R \sin{\theta/2}$ where $R$ is the radius of the circle.  The double integral may then be written
\begin{equation}
(2\pi R) 2 \int^{\pi}_{\epsilon/R} R d\theta \frac{\cos{\theta}}{(2R \sin{\theta/2} )^2},
\end{equation}
where $\epsilon$ is the spatial cutoff (so that $\epsilon/R$ is the smallest angle available) and where I have kept only the $\delta^{i j}$ piece of the $\phi \phi$ correlator.  This integral is divergent as $\epsilon \rightarrow 0$ as the entanglement entropy should be and for finite $\epsilon$ behaves like
\begin{equation}
S = c_1 \frac{R}{\epsilon} + c_2 + ...
\end{equation}
with $...$ containing only positive powers of $\epsilon$.  Note that there is no logarithmic term.

Let me repeat the same calculation for a spherical region of radius $R$ in $d=3$ dimensions.  I drop all unnecessary constant factors, the reader can supply these if they wish  The integral has a very similar structure with $\theta$ now a polar angle
\begin{equation}
S \sim R^2 \int^{\pi}_{\epsilon/R} R^2 d\theta \sin{\theta} \frac{\cos{\theta}}{(2 R \sin{\theta/2})^4} .
\end{equation}
I still recover the divergent boundary law term going like $R^2/\epsilon^2$, but now there is a pleasant surprise, namely the presence of a logarithmic term $\log{R/\epsilon}$.  Precisely such a universal logarithmic correction has been found in conformal field theories in odd spatial dimensions.  One can verify that our ansatz for the twist fields reproduces the pattern of logarithmic corrections for smooth regions in odd spatial dimensions \cite{gw_anom,ee_anom_1,ee_anom_2}.  Since the coefficient of this term is related to the derivative of $\alpha_n^2$, I expect that this derivative at $n=1$ is a universal quantity counting the number of degrees of freedom in the theory.  Of course, in a more general formulation we would expect $n$ dependence not just from $\alpha_n$ but also from the nontrivial higher point correlation functions of $\phi$.

Along the same lines, one can check that the ansatz above predicts that regions with sharp corners in any dimension have universal logarithmic corrections in their entanglement entropy related to the deficit angle at the corner.  The integral for a semi-infinite V shaped region of opening angle $\pi -\theta$ can be done exactly.  If I keep only the $\delta^{i j} $ term in the correlation function (the other term doesn't change the qualitative structure), then the coefficient of the logarithmic term is given by
\begin{equation}
\left(1- \frac{\theta \cos{\theta}}{\sin{\theta}}\right),
\end{equation}
with $-\pi < \theta < \pi$ and where the overall coefficient is undetermined (it is related to $\partial_n \alpha_n^2 |_{n=1}$).  This result has a quadratic zero at $\theta =0$ and a linear divergence at $\theta = \pi$ in agreement with previous results in a variety of systems \cite{eecorner}.  In fact, this formula even does well in a semi-quantitative comparison with previous results provided the normalization is fixed appropriately.  To understand the quadratic zero, consider the situation where the region $A$ under study and its complement form a pure state.  Then $S_A = S_{\bar{A}}$, but if $A$ has a sharp corner with angle $\theta$ then $\bar{A}$ has a sharp corner with angle $-\theta$, and hence the coefficient of the logarithmic term must be even in $\theta$ in this case.

\subsection{Mutual information from twist fields}
Still, everything thus far is in some sense a warmup, especially since the entanglement entropy contains non-universal cutoff dependence.  To study the mutual information in this twist operator formalism additional regions must be introduced.  The mutual information between two regions $\mathcal{R}_1$ and $\mathcal{R}_2$ is related to the twist fields $K_n[\mathcal{R}_1]$ and $K_n[\mathcal{R}_2]$ via
\begin{equation}
\mathcal{I} = - \partial_n \left(\langle K_n[\mathcal{R}_1] \rangle \langle K_n[\mathcal{R}_2] \rangle - \langle K_n[\mathcal{R}_1]  K_n[\mathcal{R}_2] \rangle  \right)_{n=1}.
\end{equation}
In other words, one needs the $n$ derivative of the connected correlation function between $ K_n[\mathcal{R}_1]$ and $ K_n[\mathcal{R}_2]$.

The connected correlation function ensures that after the $n$ derivative the ``self entanglement" terms $\partial \mathcal{R}_1 \partial \mathcal{R}_1$ and $\partial \mathcal{R}_2 \partial \mathcal{R}_2$ cancel.  One is left only with an integral over $\partial \mathcal{R}_1 \partial \mathcal{R}_2$ that is not divergent so long as the two regions do not touch.  Of course, I was originally interested in the singularities that develop in the mutual information precisely when the regions are brought close together.  Let me first consider the standard case of two colliding flat strips in $d=2$ dimensions.  Following the calculations above, the mutual information takes the schematic form
\begin{equation}
\mathcal{I} \sim \int^{L/2}_{-L/2} dy_1 \int^{L/2}_{-L/2} dy_2 \frac{1}{(y_1-y_2)^2 + x^2},
\end{equation}
where as before the $y$ coordinates run along the length of the strips and $x$ is the separation between the strips with $L \gg x$.  Note that in this case the product of the surface normals is independent of $y_1$ and $y_2$.  The integral is done by switching to center of mass and relative coordinates with the now familiar result $\mathcal{I} \sim L/x$.

The cases of points with finite radius of curvature and corners can be treated in a similar manner, and I obtain the scaling forms described in detail above.  Thus my ansatz for the twist fields reproduces the singularity structure of the mutual information in any dimension and in any of the collision scenarios considered above.  It also naturally accounts for the divergent structure of the bare entanglement entropy including the presence of various kinds of universal logarithmic terms.  Although I have focused on the entanglement entropy and the mutual information, the ansatz above predicts very similar behavior for the more general Renyi entropy and Renyi mutual information.  It is also possible, by introducing a length scale into the $\phi^i \phi^j $ correlator, to see the cross over structure of the entanglement entropy and mutual information in a theory with a finite correlation length. Of course, I have not derived this prescription from any particular conformal field theory, but this is a very tempting target for future calculation, especially in free conformal field theories such as the Lifshitz theory or free Dirac fermions.

I wish to mention one unsatisfactory feature of the discussion above.  The mutual information should bound the square of the connected correlation functions between any two local operators.  Considering the limit where the regions $\mathcal{R}_1$ and $\mathcal{R}_1$ are far apart, the assumption of Gaussian correlations gives a decay for the mutual information going like $1/x^{2d}$.  For free Dirac fermions, for example, this does indeed exactly bound the square of the free fermion correlation function.  However, for massless bosons, the square of the equal time correlation function decays as $1/x^{2(d-1)}$ and the bound appears to fail in our setup.  This is not totally unexpected.  In $1+1$ dimensions the boson correlation function actually grows logarithmically, a behavior clearly not in line with the decay of the mutual information.

One way out of this issue for lattice bosons is the observation that the operator norm of the lattice boson field $q_i \sim (a_i + a_i^+)$ actually diverges, or in other words, $q_i$ is an unbounded operator.  If the operator norm is infinite then the bound provided by the mutual information is vacuous.  On the other hand, there do exist systems, for example certain magnetically ordered spin systems, where the relevant operators have bounded norm and the correlation function still decays like a free boson.  What this appears to be telling us is that the long distance properties of the twist operators are considerably more variable than is captured in our assumption of Gaussian correlations.  In particular, while the second cumulant may suffice for capturing short distance singularities, one must consider higher order cumulants when evaluating $\langle K_n \rangle$ in order to correctly capture the long distance behavior.  Consider a simple example.  The mutual information of free fermions in one dimension decays as $1/x^2$ precisely in line with the fermion-fermion correlation function.  However, this system is equivalent to an XX ferromagnetic spin chain via a Jordan-Wigner transformation.  In the spin formalism we find that the spin-spin correlation function decays as $x^{-1/2}$.  This decay does not violate the bound from mutual information in the fermion language since the spin operator is non-local in the fermion language.

\subsection{A simple example: Dirac fermions}
There remains the possibility that the story I have sketched above may be close to exact for a particular system, and a good candidate for that system seems to be Dirac fermions in any dimension.  We can argue as follows.  Consider a set of $n$ replica fields $\psi_n$ each corresponding to a free Dirac fermion.  I work in $3+1$ dimensions for concreteness.  These fields carry a representation of the symmetric group $S_n$ and hence a representation of the $Z_n$ subgroup generated by the twist $\mathcal{T}^{\pm}: \psi_n \rightarrow \psi_{n\pm 1}$.  One can make a unitary transformation to a new set of fields $\psi_q$ that are eigenstates of $\mathcal{T}^{\pm}$ with eigenvalue $\lambda_q$.  These fields will pick up a phase shift as they encircle the twist operators $K_n$ in spacetime (remember, this makes sense because the twist fields live in a codimension $2$ locus in spacetime) in a fashion reminiscent of the Aharonov-Bohm effect.  Indeed, since $\lambda_q$ satisfies $|\lambda_q | = 1$, one can introduce gauge fields $A_q$ that couple to the $\psi_q$ to implement the phase $\lambda_q$.  These gauge fields are pure gauge everywhere except along the locus of definition of the twist operator.  In $3+1$ dimensions, this locus is a closed two dimensional spatial surface that is a spatial analogue of the more familiar spacetime worldsheet of a flux tube or solenoid loop in $3+1$ dimensions.  A similar approach has been used for Dirac fermions in $1+1$ dimensions \cite{ee1df}.

Because the $Z_n$ subgroup acts on the $\psi_q$ just like the global $U(1)$ charge symmetry, I can use the $U(1)$ current to couple to the gauge fields $A_q$.  Indeed, the Lagrangian in the terms of $\psi_q$ is identical in form to the Lagrangian in terms of the $\psi_n$ since the theory is free and the transformation from $\psi_n$ to $\psi_q$ is unitary.  The twist operators thus take the schematic form
\begin{equation}
K_n[\mathcal{R}] \sim \exp{\left(i \sum_q \int A_q \cdot J_q\right) },
\end{equation}
where the $A_q$ depend on $\partial \mathcal{R}$ and encode the flux needed to produce a phase shift of $\lambda_q$.  Now introduce a new field by writing $J^{\mu} = \epsilon^{\mu \nu \lambda \sigma} \partial_{\nu} \phi_{\lambda \sigma}$ for each current $J_q$.  Integrating by parts produces an integral of a $2$-form field $\phi_{\mu \nu}$ over the surface $\partial \mathcal{R}$ exactly as above.  Since the surface is purely spatial, the spacetime $2$-form descends to a spatial $2$-form and can be converted to a spatial vector using the spatial $3$-metric.  This is the setup described above.  For example, the scaling dimension of $J$ is $3$ and thus the dimension of $\phi_{\mu \nu}$ is $2$.  Also, $\phi_{\mu \nu}$ by definition has the gauge freedom I mentioned above.  However, unlike in $1+1$ dimensions, the field $\phi_{\mu \nu}$ is not Gaussian, so the story is not as simple.  There are also additional subtleties associated with fermion minus signs.  I leave to future work more detailed calculations in this case.

\section{Possibility of a generalized c-theorem}
Having investigated in some detail the singularity structure of the mutual information in various settings, let me now turn to a concrete potential application of the results described above.  There is the interesting possibility of identifying quantities in quantum field theory that are monotonic under a renormalization group flow.  Of course, such quantities need not exist in general.  However, the c-theorem of $1+1$ dimensional conformal field theory states that there is such a quantity in $1+1$ dimensions.  The quantity is the central charge of the conformal field theory, which can be defined via the two point function of the stress-energy tensor.  This quantity is guaranteed to be monotone under RG flow in any unitary conformal field theory in $1+1$ dimensions.  Remarkably, it is precisely this quantity which controls the size of the universal logarithmic divergence in the mutual information in $1+1$ dimensions.  Thus we may phrase the $1+1$ c-theorem in the following way: for any two CFTs $\mathcal{C}_1$ and $\mathcal{C}_2$ such that $\mathcal{C}_1$ flows to $\mathcal{C}_2$ under some relevant deformation, the ``entanglement per scale" $k$ as encoded in the short distance divergence of the mutual information, satisfies $k_2 \leq k_1$.  But now this opens the possibility that this result could be true for CFTs in any dimension.

There is some evidence for this claim beyond the $1+1$ dimensional setting.  For example, it is known that holographic theories with a bulk consisting of Einstein gravity coupled to matter satisfying the null energy condition possess an analogous monotone quantity \cite{holo_c_1,holo_c_2,holo_c_3}.  Moreover, this quantity is also interpretable as the number of degrees of freedom in the dual field theory, and it is, up to irrelevant numerical factors, precisely what appears in front of various universal divergences in the mutual information.  Of course, this is not the first quantity that has been proposed to satisfy a higher dimensional c-theorem.  Previous work has focused on the anomaly coefficients $a$ and $c$ in $3+1$ dimensions which control the trace of the stress-energy tensor in curved backgrounds.  In the holographic setup, these quantities are known to be all related to my proposal in terms of mutual information, at least in $1+1$ and $3+1$ dimensions.  One advantage of the mutual information based proposal is that it applies also in even spatial dimensions for which the usual anomalies are absent.

Still, there are counter-examples showing that both $a$ and $c$ are not monotone under RG flow in $3+1$ dimensions.  Perhaps the coefficient of the divergence of the mutual information can also be shown to lack monotonicity.  If one could show that the universal divergences considered here are related to $a$ and $c$ in $3+1$ dimensions, then one could conclude that the ``entanglement per scale" is not monotone in general.  Such a relation might be expected to exist since the twist operators are associated with conical singularities in spacetime, but the standard anomaly calculations must be modified in the presence of multiple disjoint regions.

There is an even more immediate objection to this proposal, namely the possibility of flowing outside the class of relativistic CFTs.  For example, by perturbing the $3+1$ dimensional Dirac fermion CFT with a finite chemical potential, one flows to a new non-relativistic fixed point in the Fermi liquid class.  Both such theories do have universal divergences in mutual information, but now even the most basic form of the scaling relation is different.  It is true that the low energy theory has, in a sense, a much weaker singularity than the high energy theory.  Perhaps this is the form of the mutual information c-theorem in this case, but we do not yet know if this can be made completely unambiguous.  An interesting holographic version of this flow comes perturbing a holographic CFT$_{3+1}$ by a finite chemical potential for some conserved U($1$).  This is described on the gravity side as an extremal black hole in AdS$_{4+1}$ with a near horizon AdS$_{1+1}$ region dual to a low energy $0+1$ dimensional conformal field theory.  The low energy $0+1$ dimensional CFT has finite entropy due to a local ground state degeneracy but no entanglement as measured by the holographic mutual information.  Thus the scaling form of the mutual information is also changed and again to a weaker sort of singularity (none at all).  The possibility of an renormalization group monotone is intriguing and deserves further study.

\section{Conclusions}
In this work I have studied the properties of mutual information in various kinds conformal field theories and scale invariant theories.  I argued that the mutual information is a cutoff independent version of the entanglement entropy which encodes the same universal physics and more.  In particular, the area law and various universal logarithmic corrections to the entanglement entropy are also manifest in the mutual information.  The original purpose of this work was to develop and clarify the concept of ``entanglement per scale" that appears clearly in $1+1$ dimensional CFT \cite{eeqft,geo_ent}, in various tensor network approaches to critical systems like MERA \cite{mera}, and in holographic duality \cite{holoreview}.  I have shown that this notion can be extended in a meaningful and precise way to all conformal field theories in any dimension and to other scale invariant theories.

With the concept of ``entanglement per scale" firmly in hand, one can begin to ask about applications of this idea.  One possibility is a classification of conformal field theories in terms of entanglement.  There has already been great progress from this point of view for gapped phases in $1+1$ and $2+1$ dimensions. More generally, the question of just how much information about the system is encoded in the ground state wavefunction deserves more systematic exploration.  It is possible that the coefficient of the universal divergence in the mutual information provides a more or less unique labeling of conformal field theories.  However, this is not quite the case even in $1+1$ dimensions.  For example, a free compactified boson always has $c=1$, but the radius of the boson is a marginal operator in the theory, thus there are many CFTs which have the same universal divergence in mutual information.  On the other hand, as the classification of minimal models with $c<1$ shows, the mutual information can be used to label $1+1$ dimensional CFTs in some situations.

One possible way forward on this question is to consider in more detail the properties of the higher dimensional twist operators considered above.  There is some hint that these operators, which encode the entanglement structure, can be classified from an algebraic point of view, but one must first understand much more about their properties.  It would be desirable to study some concrete realizations of these higher dimensional twist operators and to better understand the range of possibilities once the restrictive assumption of Gaussian correlations is relaxed.  Along these same lines, the question of monotonicity of the mutual information should also be studied in more detail.

Another virtue of the twist operator formalism I have proposed here is the relative ease of calculation involved.  Part of the interest in the subject of holographic entanglement entropy has arisen because of the particularly simple and transparent calculational structure.  Here I have argued that this structure, which is also visible in tensor network approaches like MERA for critical systems, is quite general and is a simple manifestation of the basic renormalization group structure of all local quantum field theories.  Thus at this level, I believe holographic duality accurately captures much of the structure of many body entanglement that has so far been identified.  I hope that the twist operator formalism will, as in the one dimensional case, greatly simplify calculations of entanglement entropy and mutual information higher dimensions.

I have studied the mutual information in conformal field theories in general dimension in the hopes of learning about the general structure of entanglement in quantum field theory.  Such a study is relevant for understanding the low energy structure of entanglement in general many body systems which flow to continuum quantum field theories in the infrared.  I hope that the ideas outlined here will prove useful for further explorations into the structure of many body entanglement so that we may one day have a complete theory of this mysterious substance from which quantum phases are built.

I thank Xiao-Gang Wen for support during this project, and I thank Xiao-Gang and John McGreevy for many delightful conversations about entanglement.  I also thank Maissam Barkeshli for helpful comments on the draft.  Finally, I thank a power outage on Bailey Island for inspiration.

\bibliography{mi_divergence}

\end{document}